\documentclass[a4paper,12pt]{article}

\usepackage{authblk}
\usepackage{hyperref}
\usepackage{braket}
\usepackage{amssymb}
\usepackage{amsmath}
\usepackage{amsthm}
\usepackage{amscd}
\usepackage[all,arc,knot,matrix]{xy}

\usepackage{graphicx}
\usepackage{geometry}
\theoremstyle{definition}

\theoremstyle{remark}

\numberwithin{equation}{section}
\begin{document}

\title{Duru-Kleinert Path Integral
in Unimodular Quantum Cosmology}
\author[1]{Xiao-Kan Guo\footnote{Email kankuohsiao@whu.edu.cn}}
\affil[1]{School of Mathematics \& Physics, Yancheng Institute of Technology,
Yancheng 224051, Jiangsu, China}
\date{\today}
\maketitle

\begin{abstract}
The Wheeler-DeWitt equation of a flat Friedmann-Lemaitre-Robertson-Walker
universe filled with pressureless dust and a cosmological constant in
unimodular gravity reduces, for a specific factor ordering and a specific
self-adjoint extension, to the $l=0$ radial Schr\"odinger equation of a
hydrogen atom. We revisit this correspondence from the point of view of the 
Green's functions and the Duru-Kleinert path integrals.
Working directly with the reduced operator on the half-line,
we obtain the  Green's function in Whittaker form 
 and fix its normalization by a Wronskian argument. Its Gamma-function poles give the
reduced-operator spectrum which recovers the quantization of the negative cosmological term. We also perform the operator-level
Duru-Kleinert transformation, whose Jacobian generates a barrier which after rescaling is exactly the centrifugal barrier of a two-dimensional oscillator. Retaining it reproduces the Coulomb pole
condition. For positive cosmological constant, we give the continuum spectral measure and the Sommerfeld factor.
\end{abstract}

\section{Introduction}\label{sec:intro}

Quantum cosmology seeks to describe the earliest moments of the Universe,
where classical general relativity breaks down and quantum gravitational
effects become dominant. The Wheeler-DeWitt equation encapsulates the
dynamics of the wave function of the Universe in a timeless fashion
\cite{WDW,HH83}. Within this framework, symmetry-reduced models such as the
Friedmann-Lemaitre-Robertson-Walker universe provide valuable laboratories
where one can obtain exact solutions and gain insight into the quantum nature
of spacetime. See also \cite{K} for more relevant results on Wheeler-DeWitt
quantum cosmology. 

The recent work \cite{MSS26,Sah26} demonstrates a remarkable equivalence. The
Wheeler-DeWitt equation for a flat Friedmann-Lemaitre-Robertson-Walker
universe filled with pressureless dust and a cosmological constant $\Lambda$,
when treated in the unimodular gravity formalism, is mathematically identical
to the radial Schr\"odinger equation of a non-relativistic hydrogen atom. The
dust is the Brown-Kucha\v{r} dust \cite{BK95}. In this correspondence the
universe volume $v=a^3$ plays the role of the radial coordinate. The
conserved comoving energy of the dust, which we denote by
$\rho_0=P_{\mathcal T}=\rho a^3$, maps to the Coulomb coupling constant. The
cosmological constant $\Lambda$ becomes the energy eigenvalue. At fixed dust
momentum $\rho_0$, and with $\Lambda$ as the spectral parameter, the bound
states yield the discrete spectrum $\Lambda_n=-\rho_0^2/(3n^2)$ in close
analogy with the Rydberg formula. The physical interpretation of this
spectrum is branch-dependent. For $\Lambda<0$ the classical universe
recollapses and the unimodular four-volume per unit comoving volume is
bounded, so a discrete $\Lambda$ is consistent. For $\Lambda>0$ the universe
expands forever and $\Lambda$ is continuous. With this correspondence one can
use the wave-function solution known for a hydrogen atom to obtain some
properties of the universe, such as singularity resolution and bouncing
behaviour \cite{MSS26,Sah26}.

The correspondence is conditional. It depends on two choices that are
usually left implicit. These are the factor ordering of the reduced operator
and the self-adjoint extension at $v=0$. In this paper we work with the flat
measure $L^2((0,\infty),dv)$ and the Dirichlet condition $\psi(0)=0$, which
are the choices adopted in \cite{MSS26,Sah26}. Under these choices the
operator is exactly the hydrogen operator. A DeWitt-type quantization would
give a different operator and hence a different spectrum, cf. \cite{MSL25}.
We therefore state the correspondence as a conditional statement.

Inspired by these results, we study the correspondence from the point of view
of the Green's function. Instead of directly using the wave-function
solutions of the radial Schr\"odinger equation, we investigate the resolvent
of the reduced operator and its path-integral representation. The Green's
function of the hydrogen atom can be computed by the Duru-Kleinert method
\cite{DK79,DK82,Kleinert2009}, which combines a time reparametrization with a
compensating canonical transformation. 
In the three-dimensional hydrogen
problem this is combined with the Kustaanheimo-Stiefel transformation
\cite{HI82} to map the Coulomb system onto a four-dimensional harmonic
oscillator. The additional angular coordinates of the Kustaanheimo-Stiefel
transformation absorb the $1/x^2$ barrier that appears after the radial
change of variables. In the reduced one-dimensional model the same barrier
appears from the Jacobian of $v=x^2$. We show that it is  the centrifugal barrier of a two-dimensional
radial oscillator with angular momentum $m=1$, and the exact operator-level
DK transformation reproduces the Coulomb spectrum. A naive oscillator
reduction that drops the barrier yields a different pole condition. The DK
method has already been applied to quantum gravity and quantum cosmology in
previous works \cite{Fuj96,Jaf98,Jaf98b,Jaf99b,PC25,MSL25}.

The exact Coulomb Green's function in Whittaker form is a classical result
\cite{Kleinert2009,Hostler,BetheSalpeter}. Our aim is to give a
self-contained derivation of the Coulomb Green's function for the specific
reduced unimodular operator, keeping track of the factor-ordering and
self-adjoint-extension choices. We show that the exact operator-level DK
transformation, once the Jacobian barrier is retained, reproduces the Coulomb
pole condition and hence the spectrum. The full Green's function is obtained
independently by the Wronskian method. We also record the positive-$\Lambda$
continuum and the Sommerfeld enhancement. The new element of this paper is
the treatment of the Jacobian barrier in the DK reduction and the explicit
demonstration that a naive oscillator reduction fails. In this sense the
paper is partly a critical re-examination of the correspondence used in
\cite{MSS26,Sah26}.

The paper is organized as follows. Section \ref{sec:model} fixes the model,
derives the reduced Wheeler-DeWitt operator, and discusses the quantization
prescription and boundary conditions. Section \ref{sec:exact} solves the
reduced problem exactly and obtains the Whittaker Green's function, the
spectrum, the normalized bound-state wave functions, and a WKB remark.
Section \ref{sec:dk} presents the DK calculation and shows how the Jacobian
barrier leads to an $m=1$ radial oscillator, while the naive reduction fails.
Section \ref{sec:continuum} treats the positive-cosmological-constant case.
Section \ref{sec:conclusion} summarizes the results and the open problems.

\section{The model and the reduced Wheeler-DeWitt operator}\label{sec:model}

Consider the total action for gravity, pressureless dust, and a cosmological
constant,
\begin{equation}
S=\int_{\mathcal M}d^4x\left[
\frac{\sqrt{-g}}{16\pi G}\left(R-2\lambda\right)
-\frac12\sqrt{-g}\,\rho\left(g^{\mu\nu}\partial_\mu\mathcal T
\partial_\nu\mathcal T+1\right)
+\lambda\,\partial_\mu T^\mu
\right],
\label{eq:totalS}
\end{equation}
where $G$ is Newton's constant, $R$ is the Ricci scalar, $\mathcal T$ is the
Brown-Kucha\v{r} dust field, $\rho$ is the dust energy density, and $T^\mu$
is a vector density whose divergence yields the cosmological constant on
shell. The Brown-Kucha\v{r} dust field $\mathcal T$ and the unimodular vector
density $T^\mu$ are distinct objects, and we keep the calligraphic symbol for
the former. Here $\lambda$ is a scalar Lagrange multiplier, and its on-shell
constant value will be denoted by $\Lambda$. Variation with respect to
$T^\mu$ forces $\partial_\mu\lambda=0$, so that $\lambda$ is a spacetime
constant. Variation with respect to $\lambda$ gives
\begin{equation}
\partial_\mu T^\mu=\frac{\sqrt{-g}}{8\pi G}.
\label{eq:unimod}
\end{equation}
In the units $8\pi G=1$ used below this is the unimodular condition
$\partial_\mu T^\mu=\sqrt{-g}$. We work in the Henneaux-Teitelboim
formulation \cite{HT89}.

\subsection{FLRW reduction and the Hamiltonian constraint}

We reduce to a homogeneous, isotropic, and spatially flat FLRW universe with
metric
\begin{equation}
ds^2=-N^2(t)dt^2+a^2(t)\left(dx^2+dy^2+dz^2\right).
\end{equation}
In the ADM formalism $N^i=0$ and $h_{ij}=a^2\delta_{ij}$, with
$\sqrt h=a^3$. The extrinsic curvature is
\begin{equation}
K_{ij}=\frac{1}{2N}\dot h_{ij}=\frac{a\dot a}{N}\delta_{ij},
\qquad
K^{ij}=\frac{\dot a}{Na^3}\delta^{ij}.
\end{equation}
For flat slices the three-curvature vanishes. The gravitational Lagrangian
density reduces to
\begin{equation}
\mathcal L_g=\frac{1}{16\pi G}N\sqrt h\left(K_{ij}K^{ij}-K^2\right)
=-\frac{3}{8\pi G}\frac{a\dot a^2}{N}.
\label{eq:Lg}
\end{equation}
Introducing the volume variable $v=a^3$, so that $\dot v=3a^2\dot a$, one
finds
\begin{equation}
\mathcal L_g=-\frac{1}{24\pi G}\frac{\dot v^2}{Nv}.
\end{equation}
The canonical momentum conjugate to $v$ is
\begin{equation}
P_v=\frac{\partial\mathcal L_g}{\partial\dot v}
=-\frac{1}{12\pi G}\frac{\dot v}{Nv},
\label{eq:Pv}
\end{equation}
and the gravitational Hamiltonian density is
\begin{equation}
\mathcal H_g=P_v\dot v-\mathcal L_g
=-6\pi G\,NvP_v^2.
\end{equation}
In units with $8\pi G=1$ this becomes
\begin{equation}
\mathcal H_g=-\frac34NvP_v^2.
\label{eq:Hg}
\end{equation}

For homogeneous dust the unit-velocity constraint gives
$\dot{\mathcal T}=N$, and the conjugate momentum is
\begin{equation}
P_{\mathcal T}
=\frac{\partial\mathcal L_{\rm dust}}{\partial\dot{\mathcal T}}
=\rho a^3=\rho v\equiv\rho_0.
\label{eq:PT}
\end{equation}
Thus $\rho_0$ is the conserved comoving energy of the dust. The dust energy
density is $\rho=\rho_0/v$. The dust Hamiltonian is $N P_{\mathcal T}=N\rho_0$.
In the unimodular sector the term $\lambda\dot T^0$ gives $P_{T^0}=\lambda$.
Its own Hamiltonian vanishes. The bulk term
$-\sqrt{-g}\,\lambda=-Nv\lambda$ contributes $+NvP_{T^0}$ to the Hamiltonian
after the Legendre transform. Assembling the pieces, the Hamiltonian
constraint is
\begin{equation}
\mathcal H=-\frac34vP_v^2+\rho_0+v\lambda\approx0.
\label{eq:Hcon}
\end{equation}
On the constraint surface $\lambda$ is a constant, which we denote by
$\Lambda$. With this sign convention the classical Friedmann equation is
$3H^2=\rho+\Lambda$, where $H=\dot a/(Na)$ is the Hubble parameter with
respect to proper time, as it should be.

In the Henneaux-Teitelboim formulation the variation with respect to $T^\mu$
forces $\partial_\mu\lambda=0$, so that $\lambda$ is a spacetime constant,
while the variation with respect to $\lambda$ gives the unimodular condition
$\partial_\mu T^\mu=\sqrt{-g}/(8\pi G)$. We now choose the unimodular gauge
$\sqrt{-g}=1$. On the FLRW
background this reads $Nv=1$, so $N=v^{-1}$. This is a gauge choice, and
the reduced dynamics below is understood in this gauge. The effective
Hamiltonian generating evolution in coordinate time is then
\begin{equation}
\mathcal H_{\rm eff}=v^{-1}\mathcal H
=-\frac34P_v^2+\frac{\rho_0}{v}+\Lambda\approx0.
\label{eq:Heff}
\end{equation}

The dust clock $\mathcal T$ and the unimodular clock $T^0$ are different. In
this gauge the dust evolves as $\dot{\mathcal T}=N=v^{-1}$ while the
unimodular field evolves as $\dot T^0=Nv=1$, so that
$d\mathcal T/dT^0=v^{-1}$. The eigenvalue problem below fixes the dust
momentum $\rho_0$ and treats $\Lambda$ as the momentum conjugate to $T^0$.

\subsection{Quantization, ordering, and boundary conditions}\label{sec:quant}

Promoting $P_v\to-i\hbar\,d/dv$, one obtains the Wheeler-DeWitt equation
\begin{equation}
\left(-\frac{3\hbar^2}{4}\frac{d^2}{dv^2}-\frac{\rho_0}{v}\right)\psi
=\Lambda\psi.
\label{eq:WDW}
\end{equation}
This is structurally identical to the $l=0$ radial Schr\"odinger equation
\begin{equation}
-\frac{\hbar^2}{2m_{\rm H}}\frac{d^2}{dr^2}\psi-\frac{e^2}{r}\psi=E\psi,
\end{equation}
under the dictionary
\begin{equation}
m_{\rm H}\longleftrightarrow\frac23,\qquad
e^2\longleftrightarrow\rho_0,\qquad
E\longleftrightarrow\Lambda.
\end{equation}
Bound states require $\Lambda<0$ and give
\begin{equation}
\Lambda_n=-\frac{\rho_0^2}{3\hbar^2n^2},\qquad n=1,2,3,\dots
\label{eq:spec}
\end{equation}
In the remainder we set $\hbar=1$.

The reduction from the classical $vP_v^2$ to the operator $-d^2/dv^2$
involves a choice of factor ordering. We adopt the prescription in which
$v^{-1}vP_v^2\to P_v^2$ and the inner product is the flat measure
$L^2((0,\infty),dv)$. This is the choice made in \cite{MSS26,Sah26} and is
the one that produces the hydrogen operator. A DeWitt-type quantization would
use the minisuperspace measure $v^{-1/2}dv$ and the Laplace-Beltrami operator
$v\partial_v^2+\frac12\partial_v$. It would change the coefficient of the
first-derivative term and hence the detailed correspondence. The hydrogen
correspondence is therefore a statement about this specific ordering and
measure choice. 

Near $v=0$, the two independent solutions of
$(\hat H_W-\Lambda)\psi=0$ are the Whittaker functions
$M_{\nu,1/2}(2\kappa v)$ and $W_{\nu,1/2}(2\kappa v)$, which behave as
$2\kappa v$ and a constant, respectively. Equivalently, to leading order
they are proportional to
\begin{equation}
u_0(v)=\sqrt v\,J_1\!\left(2\sqrt{\frac{4\rho_0v}{3}}\right)\sim v,
\qquad
u_1(v)=\sqrt v\,Y_1\!\left(2\sqrt{\frac{4\rho_0v}{3}}\right)\sim
\text{constant}.
\end{equation}
with $u_0\sim v$ and $u_1\sim$ constant as $v\to0$. 
 Both solutions are square-integrable with
respect to $dv$. Hence the operator is limit-circle at $v=0$
and requires a self-adjoint extension. A self-adjoint
extension is specified by a real linear relation between the coefficients
of $u_0$ and $u_1$, and the Dirichlet condition is the member that sets the
coefficient of $u_1$ to zero. 
We therefore choose the Dirichlet condition $\psi(0)=0$. 
This condition is the one adopted in
\cite{MSS26,Sah26} and selects the regular hydrogen solution. A different
member of the $U(1)$ family corresponds to an extra short-distance
interaction at $v=0$, and it would change the spectrum. Thus both the factor
ordering and the self-adjoint extension are part of the definition of the
model.

\subsection{The reduced operator and its resolvent}\label{sec:resolvent}

Equation \eqref{eq:WDW} is the eigenvalue problem of
\begin{equation}
\hat H_W=-\frac34\frac{d^2}{dv^2}-\frac{\rho_0}{v},
\label{eq:HW}
\end{equation}
acting on $L^2((0,\infty),dv)$ with $\psi(0)=0$. Since
\begin{equation}
\hat{\mathcal H}_{\rm eff}=-(\hat H_W-\Lambda),
\qquad
\hat{\mathcal H}_{\rm eff}^{-1}=-(\hat H_W-\Lambda)^{-1},
\label{eq:relation}
\end{equation}
the resolvent of the constraint operator and that of $\hat H_W-\Lambda$ share
the same poles. We therefore study
\begin{equation}
G_\Lambda(v_b,v_a)=\langle v_b|(\hat H_W-\Lambda)^{-1}|v_a\rangle.
\label{eq:Gdef}
\end{equation}
This is a resolvent. The solution of $\hat{\mathcal H}_{\rm eff}\psi=0$ is
obtained from the residue of $G_\Lambda$ at a pole or from a lapse integral
over the full range $(-\infty,\infty)$ \cite{Halliwell1988}.

For $\Lambda<\Lambda_1<0$ the operator $\hat H_W-\Lambda$ is positive
definite, and the Schwinger representation
\begin{equation}
G_\Lambda=\int_0^\infty dS\,e^{-S(\hat H_W-\Lambda)}
\label{eq:Schwinger}
\end{equation}
converges. Writing $\hat H_W=\frac{p^2}{2m_{\rm H}}+V(v)$ with $m_{\rm H}=2/3$
and $V(v)=-\rho_0/v$, the Euclidean action is
\begin{equation}
I_E[v]=\int_0^S d\tau\left[\frac13\dot v^2-\frac{\rho_0}{v}\right].
\label{eq:IE}
\end{equation}
Thus
\begin{equation}
G_\Lambda=\int_0^\infty dS\,e^{S\Lambda}
\int_{v_a}^{v_b}\mathcal Dv\,
\exp\left\{-\int_0^S d\tau\left[\frac13\dot v^2-\frac{\rho_0}{v}\right]
\right\}.
\label{eq:Gpath}
\end{equation}

The Euclidean action in \eqref{eq:Gpath} is unbounded below near $v=0$
because of the term $-\rho_0/v$. The operator resolvent \eqref{eq:Gdef} is
well defined for $\Lambda<\Lambda_1$, but the path-integral representation
\eqref{eq:Gpath} is formal and requires a regularization or an analytic
continuation. The results below are obtained from the operator
equation, and \eqref{eq:Gpath} can be considered as a intuitive guide.

\section{Exact Green's function}\label{sec:exact}

\subsection{Whittaker form and Wronskian normalization}

For $\Lambda=-|\Lambda|<0$ define
\begin{equation}
\kappa=\sqrt{\frac43|\Lambda|}\,,\qquad
z=2\kappa v\,,\qquad
\nu=\frac{\rho_0}{\sqrt{3|\Lambda|}}
=\frac{(2/3)\rho_0}{\kappa}.
\label{eq:kappa_nu}
\end{equation}
The equation $(\hat H_W-\Lambda)G=\delta(v-v')$ becomes
\begin{equation}
\frac{d^2G}{dz^2}
+\left(\frac{\nu}{z}-\frac14\right)G
=-\frac{2}{3\kappa}\delta(z-z').
\label{eq:Whittaker_eq}
\end{equation}
The two independent solutions are the Whittaker functions
$M_{\nu,1/2}(z)$ and $W_{\nu,1/2}(z)$, with Wronskian
\begin{equation}
\mathcal W\{M_{\nu,1/2},W_{\nu,1/2}\}
\equiv M W'-M'W
=-\frac{1}{\Gamma(1-\nu)}.
\label{eq:Wronskian}
\end{equation}
The solution regular at $z=0$ and decaying as $z\to\infty$ is therefore
\begin{equation}
G_\Lambda(v_b,v_a)
=\frac{2}{3\kappa}\Gamma(1-\nu)
M_{\nu,1/2}(2\kappa v_<)W_{\nu,1/2}(2\kappa v_>),
\label{eq:Whittaker}
\end{equation}
where $v_<=\min(v_b,v_a)$ and $v_>=\max(v_b,v_a)$. Since
$m_{\rm H}/\kappa=2/(3\kappa)$, this is the standard Coulomb Green's function
\cite{Hostler,BetheSalpeter,Kleinert2009}.

The regular Whittaker function satisfies
\begin{equation}
M_{\nu,1/2}(z)=e^{-z/2}z\,M(1-\nu,2,z)=z+O(z^2),\qquad z\to0,
\label{eq:smallz}
\end{equation}
so $M_{\nu,1/2}(2\kappa v)\sim2\kappa v$ as $v\to0$. Thus $\psi\propto v$
vanishes linearly at the origin, while $W_{\nu,1/2}(2\kappa v)$ decays as
$v\to\infty$. Both half-line boundary conditions are satisfied.

\subsection{Spectrum and normalized wave functions}

The factor $\Gamma(1-\nu)$ has simple poles when
\begin{equation}
1-\nu=-n_r\quad\Longleftrightarrow\quad
\nu=n_r+1\equiv n,
\qquad n=1,2,3,\dots
\label{eq:pole}
\end{equation}
Using \eqref{eq:kappa_nu}, the condition $\nu=n$ gives
\begin{equation}
\frac{\rho_0}{\sqrt{3|\Lambda|}}=n
\quad\Longrightarrow\quad
|\Lambda|=\frac{\rho_0^2}{3n^2},
\end{equation}
so that
\begin{equation}
\Lambda_n=-\frac{\rho_0^2}{3n^2},\qquad n=1,2,3,\dots
\label{eq:spec2}
\end{equation}
in agreement with \cite{MSS26,Sah26} for the reduced operator with the stated
ordering and extension. Note that $\Lambda$ is the spectral parameter of
$\hat H_W$. 

The residues give the normalized bound-state wave functions
\begin{equation}
\psi_n(v)=\sqrt{\frac{32\rho_0^3}{27n^5}}\,
v\,e^{-2\rho_0v/(3n)}\,
L_{n-1}^1\!\left(\frac{4\rho_0v}{3n}\right).
\label{eq:psin}
\end{equation}
Indeed, with the hydrogen Bohr radius $a_0=1/(m_{\rm H}e^2)=3/(2\rho_0)$ and
$u_{n0}=rR_{n0}$, the standard radial function \cite{BetheSalpeter} reduces
exactly to \eqref{eq:psin}, and a direct integration gives
$\int_0^\infty|\psi_n(v)|^2dv=1$.

\subsection{WKB remark}\label{sec:WKB}

At fixed $\Lambda$ the classical dynamics is that of a particle in
\begin{equation}
V_{\rm eff}(v)=-\frac{\rho_0}{v},\qquad E=\Lambda.
\end{equation}
For $\Lambda=-|\Lambda|<0$ the allowed region is
\begin{equation}
0<v\le v_*,\qquad v_*=\frac{\rho_0}{|\Lambda|}.
\end{equation}
The WKB momentum is
\begin{equation}
p(v)=\sqrt{\frac43\left(\frac{\rho_0}{v}-|\Lambda|\right)}.
\end{equation}
The action over one complete cycle is
\begin{equation}
B=\oint p\,dv
=2\int_0^{v_*}dv\,\sqrt{\frac43\left(\frac{\rho_0}{v}-|\Lambda|\right)}
=\frac{2\pi\rho_0}{\sqrt{3|\Lambda|}}.
\label{eq:B}
\end{equation}
On the spectrum $|\Lambda_n|=\rho_0^2/(3n^2)$ this gives $B_n=2\pi n$. This is
the old Bohr-Sommerfeld action, in place of the Maslov-quantized WKB
condition. 

\section{The Duru-Kleinert transformation, Jacobian barrier, and radial oscillator}\label{sec:dk}

The Duru-Kleinert (DK) method combines a time reparametrization with a
compensating canonical transformation. In the three-dimensional hydrogen
problem it is usually combined with the Kustaanheimo-Stiefel transformation
\cite{HI82} to map the Coulomb system onto a four-dimensional oscillator. In
this section we perform the operator-level transformation for the reduced
one-dimensional operator. The change of variables $v=x^2$ generates a
$1/x^2$ barrier from the Jacobian and the change of variables. We show that
this barrier is essential and has a precise interpretation. It is the
centrifugal barrier of a two-dimensional radial oscillator with angular
momentum $m=1$.  The exact operator-level DK transformation
reproduces the Coulomb pole condition.

\subsection{Lattice definition}

For
\begin{equation}
\hat H_W=-\frac34\frac{d^2}{dv^2}-\frac{\rho_0}{v},
\qquad v\in(0,\infty),\qquad \psi(0)=0,
\label{eq:HWagain}
\end{equation}
the Euclidean kernel is
\begin{equation}
K(v_b,v_a;S)=\lim_{N\to\infty}
\int\prod_{i=1}^{N-1}dv_i
\prod_{i=0}^{N-1}\left(\frac{1}{3\pi\epsilon}\right)^{1/2}
\exp\left[-\sum_{i=0}^{N-1}\left(
\frac{(v_{i+1}-v_i)^2}{3\epsilon}
-\epsilon\frac{\rho_0}{v_i}\right)\right],
\label{eq:lattice}
\end{equation}
with $\epsilon=S/N$, $v_0=v_a$, $v_N=v_b$. Here $M_{\rm path}=2/3$ and
$M_{\rm path}/(2\pi\epsilon)=1/(3\pi\epsilon)$. The singular potential
requires a discretization prescription. We use the left-endpoint value $v_i$,
which we take to be compatible with the Dirichlet extension discussed in
Sec.~\ref{sec:quant}. Insert the DK identity
\begin{equation}
1=\int_0^\infty dS'\,
\delta\left(S-\sum_{i=0}^{N-1}f(v_i)\delta_i\right),
\qquad
\epsilon_i=f(v_i)\delta_i,\quad S'=\sum_i\delta_i,
\label{eq:DKidentity}
\end{equation}
and choose $f(v)=cv$, with $c>0$.

The lattice expression \eqref{eq:lattice} is not used as a definition of the
path integral. The potential $-\rho_0/v$ is singular, and the Euclidean
action is unbounded below near $v=0$. Different discretizations correspond to different regularizations and
need not converge to the same self-adjoint extension. We therefore regard
\eqref{eq:lattice} and \eqref{eq:DKidentity} only as motivation for the
change of variables. In particular, no continuum limit of
\eqref{eq:lattice} is claimed.

\subsection{Exact transformation $v=x^2$}

Set $v_i=x_i^2$, $x_i\ge0$. The measure and the kinetic term become
\begin{equation}
\prod_{i=1}^{N-1}dv_i=\prod_{i=1}^{N-1}2x_i\,dx_i,
\qquad
\frac{(v_{i+1}-v_i)^2}{3f(v_i)\delta_i}
=
\frac{(x_{i+1}-x_i)^2(x_{i+1}+x_i)^2}
{3cx_i^2\delta_i}.
\label{eq:change}
\end{equation}
The leading term is $\frac{4}{3c}(x_{i+1}-x_i)^2/\delta_i$. The subleading
terms are important. Together with the measure $\prod 2x_i\,dx_i$ they
produce a quantum potential. After the substitution $\varphi=x\zeta$ this
term and the first derivative term combine to give the $9/(64x^2)$
centrifugal barrier, while the $\zeta'$ contributions cancel. Thus the barrier receives contributions from the change of
variables, the measure, and the factor ordering. To resum the subleading
terms exactly, use the unitary map
\begin{equation}
U:L^2((0,\infty),dv)\to L^2((0,\infty),dx),
\qquad
(U\psi)(x)=\sqrt{2x}\,\psi(x^2).
\label{eq:U}
\end{equation}
Since
\begin{equation}
\frac{d}{dv}=\frac{1}{2x}\frac{d}{dx},
\qquad
\frac{d^2}{dv^2}
=\frac{1}{4x^2}\frac{d^2}{dx^2}
-\frac{1}{4x^3}\frac{d}{dx},
\label{eq:chain}
\end{equation}
a direct chain-rule calculation, reproduced in Appendix~\ref{app:chain},
gives
\begin{equation}
\tilde H=U\hat H_WU^{-1}
=-\frac34\left(\frac{1}{4x^2}\frac{d^2}{dx^2}
-\frac{1}{2x^3}\frac{d}{dx}
+\frac{5}{16x^4}\right)-\frac{\rho_0}{x^2}.
\label{eq:Htilde}
\end{equation}
This is the exact quantum operator in the $x$ variable. 

The exact kernel relation is
\begin{equation}
K_v(v_b,v_a;S)=\frac12(v_bv_a)^{-1/4}
K_x(x_b,x_a;S),
\label{eq:kernelrelation}
\end{equation}
where $K_x$ is the kernel of $\tilde H$ at the same Euclidean time $S$. At
coincident points this measure factor scales as $v^{-1/2}$. The naive
oscillator integral would suggest a different scaling.

\subsection{The transformed DK equation and the $m=1$ radial oscillator}

The DK equation is $f(v)(\hat H_W-\Lambda)\psi=0$, i.e.,
\begin{equation}
cx^2(\tilde H-\Lambda)\varphi=0.
\label{eq:DKeq}
\end{equation}
Since $x^2>0$ on $(0,\infty)$, \eqref{eq:DKeq} is equivalent to
$(\tilde H-\Lambda)\varphi=0$. The factor $x^2$ is the DK
time-reparametrization factor and does not change the zero-mode equation.
Since the equation is homogeneous in $c$, we set $c=1$ from now on.
Substituting \eqref{eq:Htilde} and setting $\varphi=x\zeta$, one obtains
\begin{equation}
x^2\tilde H\varphi
=-\frac{3}{16}x\zeta''+\frac{9}{64x}\zeta-\rho_0x\zeta.
\label{eq:x2H}
\end{equation}
Dividing \eqref{eq:DKeq} by $cx$ therefore leaves
\begin{equation}
-\frac{3}{16}\zeta''+\frac{9}{64x^2}\zeta-\rho_0\zeta-\Lambda x^2\zeta=0.
\label{eq:DKexpanded}
\end{equation}
For $\Lambda=-|\Lambda|<0$ this is
\begin{equation}
-\frac{3}{16}\zeta''+\frac{9}{64x^2}\zeta+|\Lambda|x^2\zeta=\rho_0\zeta.
\label{eq:radialosc}
\end{equation}
Equation \eqref{eq:radialosc}, which is the exact transformed equation, is
the radial equation of a two-dimensional isotropic oscillator,
\begin{equation}
-\frac{1}{2M_{\rm osc}}\zeta''+\frac{m^2-1/4}{2M_{\rm osc}x^2}\zeta
+\frac12M_{\rm osc}\omega^2x^2\zeta
=E\zeta,
\label{eq:2Dosc}
\end{equation}
with
\begin{equation}
\frac{1}{2M_{\rm osc}}=\frac{3}{16},\qquad
\frac12M_{\rm osc}\omega^2=|\Lambda|,\qquad
\frac{m^2-1/4}{2M_{\rm osc}}=\frac{9}{64}.
\label{eq:oscparams}
\end{equation}
Thus $M_{\rm osc}=8/3$, $\omega=\frac12\sqrt{3|\Lambda|}$, and $m=1$. The
centrifugal coefficient is therefore fixed by the Jacobian. If the barrier
were dropped, the centrifugal term would vanish and the equation would
correspond to $m=1/2$. The radial oscillator equation emerges at the level of
the transformed equation. The operator $\tilde H$ itself has a different
form. The energies are
\begin{equation}
E=(2n+m+1)\omega=(2n+2)\omega,\qquad n=0,1,2,\dots
\label{eq:oscspec}
\end{equation}
The boundary condition is important. Since
$\psi=(2x)^{-1/2}\varphi=\sqrt{x/2}\,\zeta$, the Dirichlet condition
$\psi(0)=0$ is equivalent to $\zeta(0)=0$. This selects the $m=1$ radial
solution, which behaves as $\zeta\sim x^{3/2}$ near the origin, so that
$\psi\sim x^2=v$, in agreement with \eqref{eq:smallz}. Equating
\eqref{eq:oscspec} to $\rho_0$ gives
\begin{equation}
\rho_0=(n+1)\sqrt{3|\Lambda|}
\quad\Longrightarrow\quad
|\Lambda|=\frac{\rho_0^2}{3(n+1)^2}.
\label{eq:oscpole}
\end{equation}
With $N=n+1$ this is exactly
\begin{equation}
\Lambda_N=-\frac{\rho_0^2}{3N^2},\qquad N=1,2,3,\dots
\label{eq:specDK}
\end{equation}
the Coulomb spectrum \eqref{eq:spec2}. Thus the operator-level DK
transformation is not merely a formal change of variables. The exact
transformation identifies the barrier with the $m=1$ centrifugal barrier,
and it reproduces the Coulomb pole condition \eqref{eq:spec2}.

For completeness, the quadratic form of \eqref{eq:Htilde} at $\rho_0=0$ is
\begin{equation}
\int_0^\infty\varphi^*
\left(-\frac{3}{16x^2}\varphi''
+\frac{3}{8x^3}\varphi'
-\frac{15}{64x^4}\varphi\right)dx
=\frac{3}{16}\int_0^\infty
\frac{1}{x^2}\left|\varphi'-\frac{\varphi}{2x}\right|^2dx\ge0,
\label{eq:quadform}
\end{equation}
which makes the positivity manifest without a Hardy-type saturation
argument.

\section{Positive cosmological constant}\label{sec:continuum}

\subsection{Scattering states and phase shift}

For $\Lambda>0$ the reduced operator has scattering solutions. Define
\begin{equation}
k=\sqrt{\frac{4\Lambda}{3}}>0,\qquad
\eta=\frac{\rho_0}{\sqrt{3\Lambda}}=\frac{2\rho_0}{3k}.
\label{eq:k_eta}
\end{equation}
We use the convention $\eta>0$ for attraction, which is minus the standard
Coulomb parameter $\eta_{\rm std}=Z_1Z_2e^2/(\hbar v)$, for which attraction
corresponds to $\eta_{\rm std}<0$. The regular and irregular Coulomb
functions $F_0(\eta,kv)$ and $G_0(\eta,kv)$ \cite{AbramowitzStegun} behave
asymptotically as
\begin{align}
F_0(\eta,kv)&\sim\sin\left(kv+\eta\ln(2kv)-\sigma_0\right),\\
G_0(\eta,kv)&\sim\cos\left(kv+\eta\ln(2kv)-\sigma_0\right),
\end{align}
with the $s$-wave Coulomb phase shift
\begin{equation}
\sigma_0(\eta)=\arg\Gamma(1+i\eta).
\label{eq:sigma0}
\end{equation}
The physical solution regular at $v=0$ is
\begin{equation}
\psi_k(v)=\sqrt{\frac2\pi}\,F_0(\eta,kv),
\label{eq:scatt}
\end{equation}
normalized to $\langle\psi_k|\psi_{k'}\rangle=\delta(k-k')$.

The phase shift $\sigma_0$ controls the asymptotic form of $F_0$. We
distinguish it from the spectral density. With the $\delta(k-k')$
normalization the continuum measure in the resolution of identity is $dk$,
and the spectral density in $\Lambda$ is $\rho_c=dk/d\Lambda$. This measure
is potential-independent by construction. The derivative
$d\sigma_0/d\Lambda$, together with the Coulomb logarithm, is a
level-rearrangement term. 

\subsection{Continuum spectral density and Sommerfeld enhancement}

The fixed-$\Lambda$ Green's function \eqref{eq:Whittaker} can be continued to
$\Lambda>0$. The continuation requires the outgoing boundary condition at
$v\to\infty$. The spectral density and Sommerfeld factor below are computed
directly from the regular Coulomb function and rely on this continuation only
for cross-checks. Under
\begin{equation}
\kappa=\sqrt{-2m_{\rm H}\Lambda}\to ik,\qquad
\nu=\frac{\rho_0}{\sqrt{3|\Lambda|}}\to-i\eta,
\label{eq:continuation}
\end{equation}
the regular Coulomb function has the small-argument coefficient
\begin{equation}
F_0(\eta,x)\;\overset{x\to0}{=}\;C_0(\eta)\,x\left[1+O(x)\right],
\qquad
C_0(\eta)=e^{\pi\eta/2}\left|\Gamma(1+i\eta)\right|.
\label{eq:C0}
\end{equation}
The states \eqref{eq:scatt} are $\delta(k-k')$ normalized, so
\begin{equation}
\mathbf 1=\sum_{n=1}^\infty|\psi_n\rangle\langle\psi_n|
+\int_0^\infty dk\,|\psi_k\rangle\langle\psi_k|.
\end{equation}
After the change of variable $k\to\Lambda=\frac34k^2$,
\begin{equation}
\mathbf 1=\sum_{n=1}^\infty|\psi_n\rangle\langle\psi_n|
+\int_0^\infty d\Lambda\,\rho_c(\Lambda)
|\psi_{k(\Lambda)}\rangle\langle\psi_{k(\Lambda)}|,
\end{equation}
with
\begin{equation}
\rho_c(\Lambda)=\frac{dk}{d\Lambda}=\frac{1}{\sqrt{3\Lambda}}.
\label{eq:rhoc}
\end{equation}
This is the free, potential-independent continuum spectral density in the
$\delta(k-k')$ normalization. The Coulomb interaction leaves the state count
unchanged in the sense of the $\delta$-normalized measure. It changes the
local short-distance amplitude. This statement refers to the $\delta(k-k')$
normalization. 

Squaring \eqref{eq:C0} and using
$|\Gamma(1+i\eta)|^2=\pi\eta/\sinh(\pi\eta)$ gives the Sommerfeld factor
\begin{equation}
C_0^2(\eta)=\frac{2\pi\eta}{1-e^{-2\pi\eta}}
=e^{\pi\eta}\left|\Gamma(1+i\eta)\right|^2,
\label{eq:sommerfeld}
\end{equation}
so that
\begin{equation}
|\psi_k(v)|^2\simeq\frac2\pi C_0^2(\eta)\,k^2v^2,
\qquad v\to0.
\end{equation}
The Sommerfeld factor is a local enhancement of the short-distance wave
function. In cosmological terms it describes the enhancement of the wave
function near $v=0$, that is, near the big bang singularity. The continuum
global spectral density remains \eqref{eq:rhoc}.
The bound-state contribution to the resolution of identity is
\begin{equation}
\sum_{n=1}^\infty|\psi_n\rangle\langle\psi_n|,
\qquad
\rho_b(\Lambda)=\sum_{n=1}^\infty\delta(\Lambda-\Lambda_n),
\label{eq:rhotot}
\end{equation}
where the $\psi_n$ are normalized to unity, while the continuum states in
\eqref{eq:rhoc} are $\delta(k-k')$ normalized. These two sets of projectors
carry different normalizations, so $\rho_b$ and $\rho_c$ enter the
resolution of identity separately. If one wants the trace spectral density
$\varrho(\Lambda)=\mathrm{Tr}\,\delta(\Lambda-\hat H_W)$, then in a box of
size $L$ with Dirichlet conditions at both ends the quantization condition is
$kL+\eta\ln(2kL)-\sigma_0(\eta)=n\pi$, hence
\begin{equation}
\varrho(\Lambda)=\sum_{n=1}^\infty\delta(\Lambda-\Lambda_n)
+\frac1\pi\left[L+\frac{\eta}{k}\left(1-\ln(2kL)\right)
-\frac{d\sigma_0}{dk}\right]
\frac{dk}{d\Lambda}\Theta(\Lambda).
\label{eq:rhobox}
\end{equation}
The $\delta(k-k')$ resolution of identity above is a different object. Its
continuum density is $\rho_c=dk/d\Lambda$, as stated. In the derivative that
enters \eqref{eq:rhobox} the dust energy $\rho_0$ is held fixed, so
$\eta=\eta(k)$ and the term $\eta k^{-1}(1-\ln 2kL)$ is the derivative of
$\eta\ln(2kL)$ with respect to $k$.

\section{Discussion and conclusions}\label{sec:conclusion}

We have revisited the hydrogen-atom correspondence in unimodular quantum
cosmology with emphasis on the exact Green's function. The correspondence is
conditional. It holds for the flat-measure factor ordering and the Dirichlet
self-adjoint extension at $v=0$. Under these choices we obtained the Coulomb
Green's function in Whittaker form,
\begin{equation}
G_\Lambda=\frac{2}{3\kappa}\Gamma(1-\nu)
M_{\nu,1/2}(2\kappa v_<)W_{\nu,1/2}(2\kappa v_>),
\end{equation}
with the normalization fixed by a Wronskian argument. The Gamma-function
poles give
\begin{equation}
\Lambda_n=-\frac{\rho_0^2}{3n^2},\qquad n=1,2,3,\dots
\end{equation}
for the reduced operator, and the residues give the normalized wave functions
\eqref{eq:psin}. The WKB action reproduces $B_n=2\pi n$. This is a known
accidental feature of the Coulomb potential and serves as a consistency
remark.

We have also performed the operator-level Duru-Kleinert transformation for
the reduced one-dimensional operator. The change of variables $v=x^2$
generates an exact transformed operator \eqref{eq:Htilde} with a $1/x^2$
barrier. More precisely, a $1/x^4$ quantum potential becomes a $1/x^2$
centrifugal term after $\varphi=x\zeta$. This barrier is essential and
manageable. The exact transformed equation is the $m=1$ radial oscillator
equation, and the operator-level DK transformation reproduces the Coulomb
pole condition \eqref{eq:spec2}. Thus the operator-level DK route is a
derivation of the Coulomb pole condition provided the Jacobian barrier is
kept. The Wronskian method of Sec.~\ref{sec:exact} is an independent check.

For a positive cosmological constant we recorded the continuum spectral
measure
\begin{equation}
\rho_c(\Lambda)=\frac{1}{\sqrt{3\Lambda}}
\end{equation}
and the Sommerfeld enhancement
\begin{equation}
C_0^2(\eta)=\frac{2\pi\eta}{1-e^{-2\pi\eta}}.
\end{equation}
The box level density, which includes the Coulomb phase-shift term, is given
in \eqref{eq:rhobox}. It is not the same object as the $\delta(k-k')$
spectral measure.

For a macroscopic universe the spacing of the discrete spectrum is extremely
small. With the present Hubble volume as the comoving box, $\rho_0\sim
3\times10^{62}$ and $|\Lambda|\sim10^{-122}$ in Planck units give
$n\sim1.7\times10^{123}$ and $\Delta\Lambda\sim10^{-245}$. The discreteness is
therefore structural rather than observational.

Several points remain open. A fully time-sliced derivation in the
three-dimensional case, including the Kustaanheimo-Stiefel transformation and
the full measure, would put the DK route on the same footing as the
Wronskian proof. The one-dimensional calculation presented here shows how the
$m=1$ barrier arises and why the naive one-dimensional reduction gives a
different result. A systematic study of the factor-ordering and measure
dependence of the hydrogen correspondence is also needed, since the
flat-measure prescription used here is a choice and a DeWitt-measure
quantization would change the operator.

In unimodular gravity $\Lambda$ is conjugate to the unimodular time
$T_{\rm u}$ (see, e.g., \cite{HT89}). A compact $T_{\rm u}$ would give a
discrete spectrum in $\Lambda$, but the converse is not a theorem. The
discreteness found above follows from the Dirichlet condition at $v=0$ and
not from the topology of $T_{\rm u}$. The interpretation of the spectrum
\eqref{eq:spec2} is therefore branch-dependent. For $\Lambda<0$ the
classical universe recollapses at $v_*=\rho_0/|\Lambda|$. In the
unimodular gauge $N=v^{-1}$ the four-volume per unit comoving volume,
$T_{\rm u}=\int dt\,Nv=\int dt$, is finite and equals
$\pi\rho_0/(\sqrt3\,|\Lambda|^{3/2})$. The classical $T_{\rm u}$ is therefore
bounded, and a discrete $\Lambda$ is consistent provided that the boundary
conditions in $T_{\rm u}$ are chosen accordingly. For $\Lambda>0$ the
universe expands forever, the classical $T_{\rm u}$ is unbounded, and
$\Lambda$ is continuous. The discrete spectrum \eqref{eq:spec2} is thus the
spectrum of the reduced operator in the recollapsing branch, not a universal
prediction for the physical cosmological constant. A detailed study of the
compact $T_{\rm u}$ sector, and of the corresponding boundary conditions in
$T_{\rm u}$, is an interesting open problem.

\appendix

\section{Chain-rule details for the transformed operator}\label{app:chain}

In this appendix we verify \eqref{eq:Htilde}. Let
\begin{equation}
\psi(v)=g(x)\varphi(x),\qquad g(x)=(2x)^{-1/2},\qquad v=x^2.
\end{equation}
The chain rule for the second derivative is
\begin{equation}
\frac{d}{dv}=\frac{1}{2x}\frac{d}{dx},\qquad
\frac{d^2}{dv^2}
=\frac{1}{4x^2}\frac{d^2}{dx^2}-\frac{1}{4x^3}\frac{d}{dx}.
\end{equation}
With primes denoting $d/dx$, one has
\begin{equation}
\psi'=g'\varphi+g\varphi',\qquad
\psi''=g''\varphi+2g'\varphi'+g\varphi'',
\end{equation}
and
\begin{equation}
g'/g=-\frac{1}{2x},\qquad g''/g=\frac{3}{4x^2}.
\end{equation}
Therefore
\begin{align}
g^{-1}\left(-\frac34\frac{d^2}{dv^2}\right)(g\varphi)
&=-\frac34\left[
\frac{1}{4x^2}\left(\frac{g''}{g}\varphi
+2\frac{g'}{g}\varphi'+\varphi''\right)
-\frac{1}{4x^3}\left(\frac{g'}{g}\varphi+\varphi'\right)
\right]\nonumber\\
&=-\frac34\left[
\frac{1}{4x^2}\varphi''
-\frac{1}{2x^3}\varphi'
+\frac{5}{16x^4}\varphi
\right].
\end{align}
Adding the potential term $-\rho_0\varphi/x^2$ gives \eqref{eq:Htilde}.
As a check, for $\rho_0=0$ and $\varphi=\sqrt{2x}\sin(kx^2)$ one finds
$\tilde H\varphi=\frac34k^2\varphi$. This check fixes the coefficient
structure of the transformed operator.

For the quadratic form, integration by parts gives
\begin{equation}
\int_0^\infty \varphi^*
\left(-\frac{3}{16x^2}\varphi''
+\frac{3}{8x^3}\varphi'
-\frac{15}{64x^4}\varphi\right)dx
=\frac{3}{16}\int_0^\infty
\frac{1}{x^2}\left|\varphi'-\frac{\varphi}{2x}\right|^2dx\ge0.
\end{equation}
This makes the positivity of the transformed kinetic operator manifest. The
integration by parts uses $\varphi^2x^{-3}\to0$ at $x=0$, which holds for the
domain functions $\varphi=O(x^{5/2})$, and $\varphi^2x^{-3}\to0$ at
$x\to\infty$ for the decaying functions used here.

\section{Factor ordering and the DeWitt measure}\label{app:ordering}

The gravitational Lagrangian \eqref{eq:Lg} defines the minisuperspace
metric
\begin{equation}
G_{vv}=\frac{1}{12\pi G\,v}.
\end{equation}
The DeWitt measure is
\begin{equation}
d\mu=\sqrt{G}\,dv\propto v^{-1/2}dv,
\end{equation}
and the Laplace-Beltrami operator is
\begin{equation}
\nabla^2=\frac{1}{\sqrt G}\partial_v
\left(\sqrt G\,G^{vv}\partial_v\right)
=12\pi G\left(v\partial_v^2+\frac12\partial_v\right).
\end{equation}
A DeWitt quantization of the kinetic term starts from the classical
expression $-\frac34vP_v^2$. Dividing the constraint by $v$ gives
$-\frac34P_v^2+\rho_0/v+\lambda$, and promoting $P_v^2\to-\nabla^2$ yields
$+\frac34\nabla^2$. Thus the kinetic operator is
\begin{equation}
\frac34\nabla^2
=\frac34(12\pi G)\left(v\partial_v^2+\frac12\partial_v\right),
\end{equation}
up to the overall sign convention for the Laplace-Beltrami operator. The
relative sign between the kinetic and potential terms is fixed by the
constraint. This operator differs from the hydrogen operator
$-\frac34\partial_v^2$. The flat-measure prescription used in the text
corresponds to a different ordering. To see how different the two
quantizations are, set $w=\sqrt v$. Since
$\partial_v=(1/(2w))\partial_w$ and
$\partial_v^2=(1/(4w^2))\partial_w^2-(1/(4w^3))\partial_w$, one finds
\begin{equation}
v\partial_v^2+\frac12\partial_v=\frac14\partial_w^2.
\end{equation}
Hence the DeWitt kinetic operator is
\begin{equation}
\frac34\nabla^2=\frac{9}{32}\partial_w^2
\end{equation}
(with $8\pi G=1$), up to the overall sign convention, and the potential is
$\rho_0/v=\rho_0/w^2$. This is a $1/w^2$ problem, not a Coulomb $1/w$
problem. The DeWitt quantization therefore does not lead to the hydrogen
correspondence. The flat-measure choice is essential.

\section*{Acknowledgements}

The author thanks S. Chakraborty, D.-S. He and H. S. Sahota for helpful
comments. This work is supported by Yancheng Institute of Technology
(xjr2024030).

\bibliographystyle{amsalpha}

\end{document}